\documentclass[aps,prl,twocolumn,superscriptaddress,groupedaddress]{revtex4}
\usepackage{graphicx}    
\usepackage{dcolumn}    
\usepackage{bm}             
\usepackage{amssymb}   
\usepackage{amsmath}    
\usepackage{subfigure}    

\begin{document}
\title{Majorana fermions in charge carrier hole quantum wires}
\author{Jingcheng Liang}
\email{liang101@purdue.edu} \affiliation{Department of Physics and Astronomy and Purdue Quantum Center, Purdue University, West Lafayette IN, 47907 USA}
\author{Yuli Lyanda-Geller}
\email{yuli@purdue.edu} \affiliation{Department of Physics and Astronomy and Purdue Quantum Center, Purdue University, West Lafayette IN, 47907 USA}

\date{March, 11 2016}
\begin{abstract}

Luttinger holes with strong Zeemann and spin-orbit interactions  in a wire proximity-coupled to a  superconductor is a promising system for observation of Majorana fermions. 
 Earlier treatments of confined Luttinger holes in wires ignored a mutual transformation of heavy and light holes upon reflection from the heteroboundaries. This effect is crucial for Zeemann and spin-orbit coupling in the ground subband of holes with several  spin-orbit terms linear in momentum. We discuss the criterion for realizing Majorana modes in charge carrier hole systems. GaAs or InSb hole wires shall exhibit stronger topological superconducting pairing compared to InSb electron systems in similar or weaker magnetic fields.  

\end{abstract}

\maketitle

Quest for Majorana zero modes is an exciting recent development in physics. Majorana particles are their own anti-particles that obey non-Abelian statistics and 
are promissing for quantum computation\cite{nayak2008non,kitaev2003fault,alicea2011non}. In condensed matter systems Majorana modes potentially arise in spinless p-wave superconductors, with candidate systems under exploration \cite{spinlessp}. Several schemes for realizing Majorana modes generate spinless p-wave superconductivity in semiconductor-superconductor hybrid structures using three ingredients: proximity effects, time reversal symmetry breaking, and spin-orbit interactions\cite{fu2008superconducting,oreg2010helical,sau2010generic,lutchyn2010majorana,lutchyn2011search,alicea2012new}.

The key problem is optimal conditions for generation and manipulation of Mayorana fermions, particularly symmetry and magnitude of spin-orbit interactions.
Strong spin-orbit coupling leads to stronger p-wave pairing \cite{alicea2010majorana}. Both electrons and hole systems were suggested for realizing Majorana modes in hybrid structures\cite{alicea2012new,mao2012hole,DasSarma2012}. However, while spin-orbit interactions are relatively well known in electron settings, their understanding in hole systems is just emerging \cite{Ivchenko,simion2014magnetic}.

Here we present the results on spin-orbit and Zeemann interactions in hole quantum wells and wires, and discuss emergence of Majorana modes in hole settings.  
Spin-orbit effects in hole wires lead to momentum-dependent Zeemann fields in all three spatial directions. Although the Bogoliubov-De Gennes (BdG) equation 
cannot have real solutions in this case, and no choice of the phase of the superconducting order parameter can decouple real and imaginary parts of the BdG wavefunctions, zero energy Majorana mode still exists. The parameter defining the ability of hybrid systems to show p-type superconductivity and Majorana modes is 
spin-orbit energy $E_{so}=\gamma^2 m$, where $m$ is the effective mass and $\gamma$ is the spin-orbit constant.
While electron $\gamma$'s are strong, $m$'s are small in InAs and InSb. Our theory shows that $E_{so}$ for holes, e.g., in InSb, can be bigger than that for electrons. 
Furthermore, the hole g-factor can exceed that for electrons, allowing for smaller magnetic fields at the same Zeeman splitting, enhancing the proximity effect. This makes 
hole systems a perspective experimental setting.

\begin{figure}
\centering
\subfigure
{
\begin{minipage}[b]{0.25\textwidth}
\includegraphics[width=4.5cm,height=2.7cm]{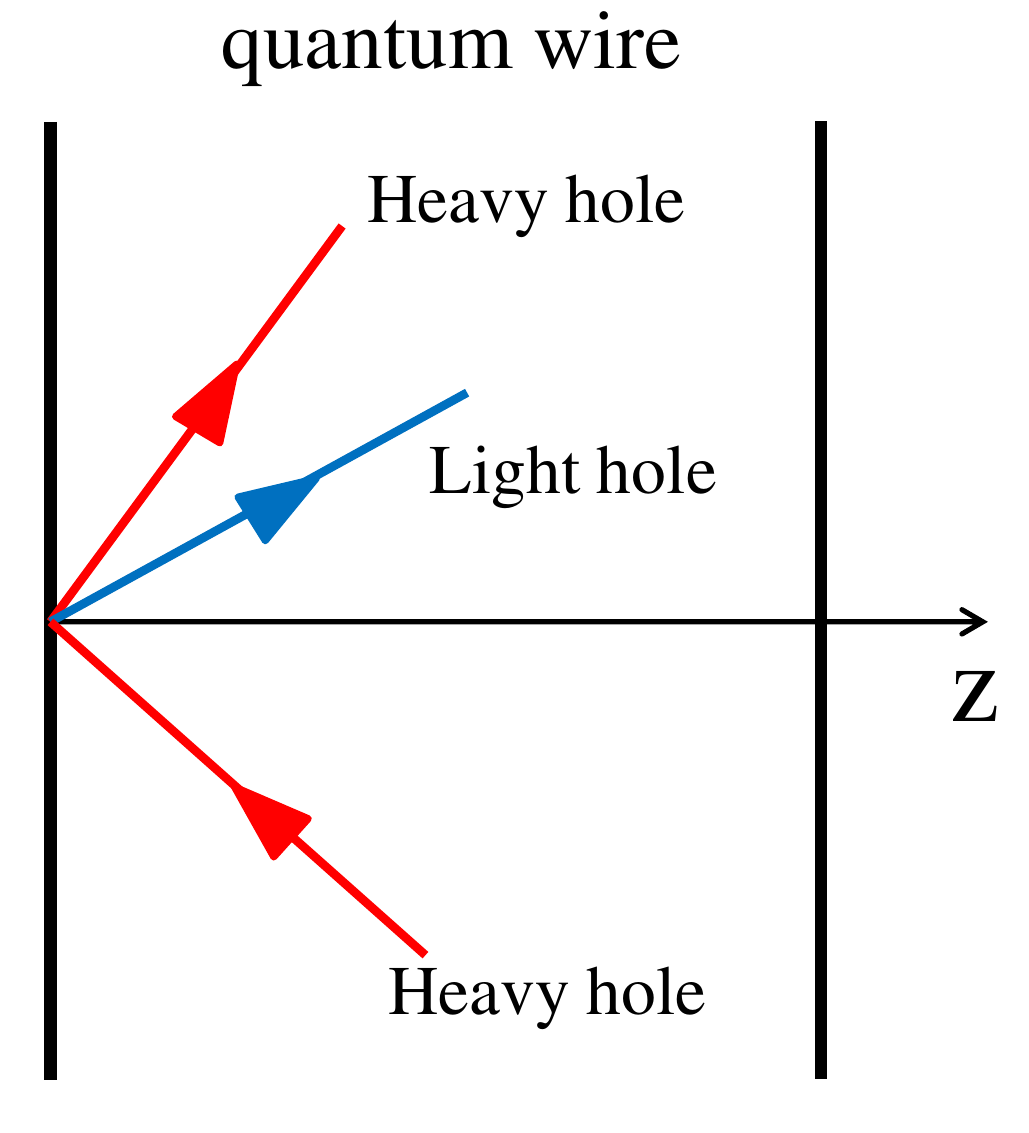}
\put(-190,65){(a)}
\end{minipage}
\label{1a}
}
\\
\subfigure
{
\begin{minipage}[b]{0.22\textwidth}
\includegraphics[width=2.5cm,height=2.2cm]{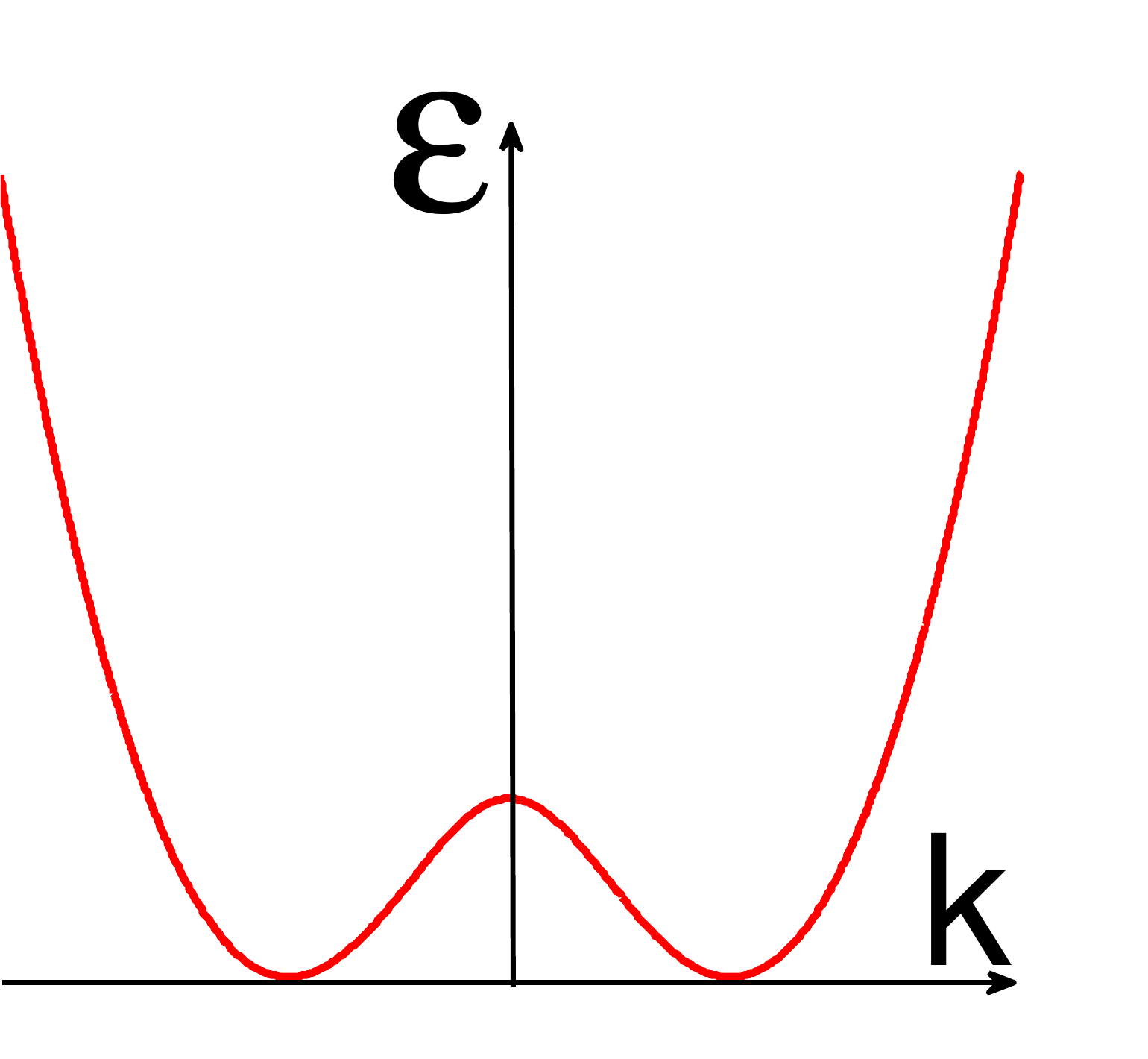}
\put(-85,60){(b)}
\end{minipage}
\label{1b}
}
\subfigure
{
\begin{minipage}[b]{0.22\textwidth}
\includegraphics[width=2.5cm,height=2.2cm]{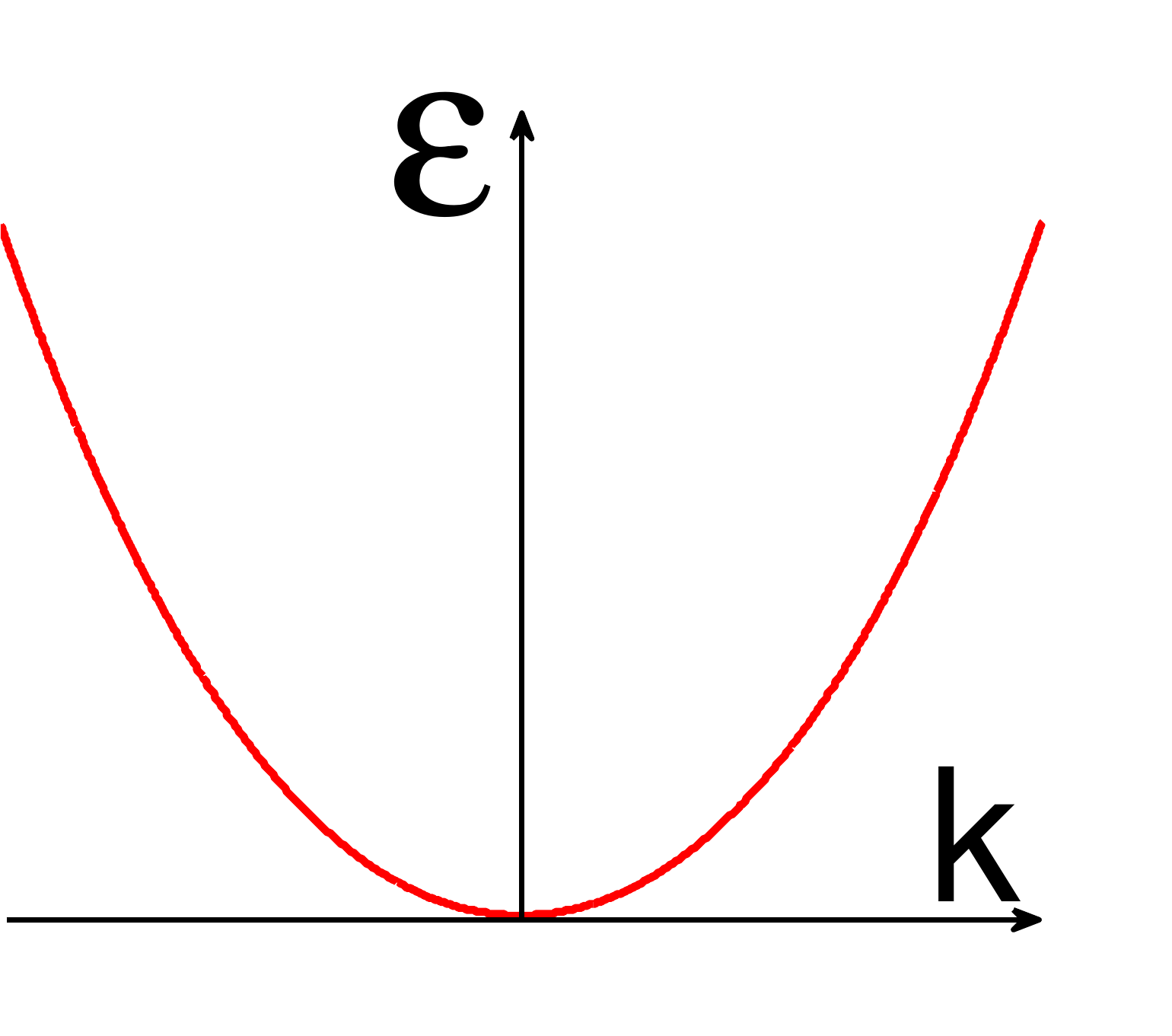}
\put(-93,60){(c)}
\end{minipage}
\label{1c}
}
\caption{(Color online). a: Transformation of holes reflected from the potential walls of the quantum wire: a heavy hole becomes heavy or light hole. b: Ground spin-orbit state energy at $E_{so}>M_z$. c: Ground state energy at $E_{so}<M_z$.  \label{fig1}}
\end{figure}

The treatment of low-dimensional holes is controversial. 
The majority of authors treats holes like electrons \cite{Arovas,LyandaGeller2004,LossPRL05,Bernevig,Hughes,mao2012hole}: if the motion of particles is quantized in direction $i$ in a well of width $d$, their Hamiltonian is solved by replacing momentum $p_i$ by zeros and $p^2_i$ by its expectation value in the ground state, $\langle \hat{ p^2_i}\rangle=(\frac{\hbar\pi}{d})^2$, and spin-orbit and Zeemann terms are found using perturbation theory. However, this approach for holes is flawed. It does not account for a mutual transformation of heavy and light holes upon reflection from the heteroboundaries,  Fig.\ref{fig1}. Although this effect can be evaluated perturbatively by including off-diagonal terms linear in $k_i$, it then requires summation  of an infinite number of terms, which are parametrically all the same\cite{simion2014magnetic}. An alternative non-perturbative approach is known since the work of Nedorezov \cite{nedorezov1971space}, but is seldom used \cite{merkulov1991momentum,rashba1988spin,lyanda2012pure,simion2014magnetic}. We show that in hole wires, this phenomenon strongly affects effective masses, g-factor and spin-orbit constants.

\paragraph{The effective Hamiltonian for hole wires.}
 We consider hole wires defined by electric gates or developed lithographically in quantum wells \cite{Rokhinson,Rokhinson2, Hamilton}, when the wire width $w\gg d$.  
The starting point is the Luttinger Hamiltonian describing the bulk holes:
\begin{equation}
\label{e1}
H_L=(A+\frac{5}{4}B)p^2-\sum_i BJ_i^2p_i^2 +D [J_iJ_{i+1}]p_ip_{i+1}],
\end{equation}
where $\mathbf{J}$ is the spin $3/2$ operator, $i=x,y,z$, $i+1=y,z,x$, $A=\gamma_1/2m_0$, $B=\gamma_2/m_0$, $D=2\gamma_3/m_0$, $\gamma_l$ are Luttinger constants, $m_0$ is the free electron mass. 

We diagonalize Eq.(\ref{e1}) in an infinite symmetric well. The symmetric and antisymmetric with respect to reflection about $z=0$  plane wavefunctions $\varphi_{+}(z,\mathbf{r})$, $\varphi_{-}(z,\mathbf{r})$, where $\mathbf{r}=(x,y)$, vanish at $z=\pm d/2$.
In the basis of Bloch functions $u^{3/2}, u^{1/2},u^{-1/2}, u^{-3/2}$ of bulk holes with $J=3/2$, $\varphi_{+,-}^{\mathbf{k}}(z,\mathbf{r})=\varphi_{+,-}^{\mathbf{k}}(z)exp(i\mathbf{k}\cdot\mathbf{r})$, and \cite{lyanda2012pure}:
\begin{equation}
\label{e2}
\varphi_+^{\mathbf{k}}(z)=
\left(\begin{array}{c}
A_0C_z\\
-iA_1S_ze^{i\phi_k}\\
A_2C_ze^{2i\phi_k}\\
-iA_3S_ze^{3i\phi_k}
\end{array}\right)
,
\varphi_-^{\mathbf{k}}(z)=
\left(\begin{array}{c}
iA_3S_ze^{-3i\phi_k}\\
A_2C_ze^{-2i\phi_k}\\
iA_1S_ze^{-i\phi_k}\\
A_0C_z
\end{array}\right),
\end{equation}
where the wavevector $\mathbf{k}\perp \hat{z}$, $\phi_k$ is the angle between $\mathbf{k}$ and $\hat{x}$, $S_z=sin(q_hz)-(s_h/s_l)sin(q_lz)$, $C_z=cos(q_hz)-(c_h/c_l)cos(q_lz)$, where $s_h=sin(q_hd/2)$, $s_l=sin(q_ld/2)$, $c_h=cos(q_hd/2)$, $c_l=cos(q_ld/2)$. For the in-plane energies much smaller than the confinement energy in the z-direction, the  heavy and light hole wave vectors in the ground state $q_h\sim\pi/d$ and $q_l=\sqrt{\nu}q_h$, $\nu=m_l/m_h$, and, in a spherical approximation ${\gamma_3}/{\gamma_2}=1$,  the coefficients  $A_0=\sqrt{d/2}$, 
$A_1=-\sqrt{3}kA_0/2q_h$, $A_2=\sqrt{3}A_0k^2/4q_h^2$,
$A_3=3A_0k^3/8\nu q_h^3$.Thus, the 2D holes are described by the two standing waves reflecting mutual transformation of heavy and light holes, Fig.\ref{fig1}.
Next we project the Hamiltonian of the system
\begin{equation}
\label{H}
H=H_L(\mathbf{K})+H_1+H_2+
U_v(z,\mathbf{r})+\tilde{M_1}_z J_z  +\tilde{M_2}_zJ_z^3
\end{equation}
on the degenerate doublet of the hole ground states $\varphi_{+,-}^{\mathbf{k}}(z,\mathbf{r})$. 
Here $\mathbf{K}=\mathbf{k}-\frac{e}{\hbar c}\mathbf{A}$,  where 
$\mathbf{A}$ is the vector-potential, $U_v(z,\mathbf{r})$ is the potential confining holes including its asymmetric part. Zeemann coupling of 2D holes to magnetic field 
$\mathbf{B}=curl \mathbf{A}$ comes from the orbital effect of magnetic field in the first term of (\ref{H}) due to a strong Luttinger $\hat{\mathbf{J}}$ to $\hat{\mathbf{p}}$ coupling, and from pure Zeemann effects, $\tilde{M_1}_z=\kappa\mu_BB_z$, $\tilde{M_2}_z=q\mu_BB_z$, where $\kappa$ and $q$ are Luttinger parameters, and $\mu_B=e\hbar/2m_0c$ is the Bohr magneton. Magnetic field is presumed applied in z-direction. 
Eq. (\ref{H}) includes the Dresselhaus terms most relevant in the hole ground state \cite{rashba1988spin,Ivchenko,simion2014magnetic}, 
\begin{equation}
H_h^{(1)}=\frac{1}{2}\delta\alpha_v
\sum_j V_jp_j\left(p_j^2-\frac{1}{3}p^2\right)-J_j^3\kappa_j, 
\label{Dres}
\end{equation}
 which give linear in $k$ terms in the 2D Hamiltonian. Here $V_z= J_xJ_zJ_x-J_yJ_zJ_y$, $\kappa_z=k_z(k_x^2-k_y^2)$, and cyclic permutation of indices $x,y,z$ defines other components of these axial vectors. We neglect linear in $J$ Dresselhaus term \cite{Pikus,Ivchenko}, which gives small  $k^3$ 2D terms and a tiny correction to 1D terms coming from  Eq.(\ref{Dres}).
Hole term $H_h^{(2)}$ in (\ref{H}) arises due to admixture of conduction electrons to holes, 
\begin{equation}
\label{Rash}
H_h^{(2)}=\zeta(\mathbf{J}\times\mathbf{p})\cdot\partial_{\mathbf{r}} U_e(z,\mathbf{r}),
\end{equation}
 where  $U_e(z,\mathbf{r})$ is the potential affecting electrons, $\zeta=-P^2/3 E_g^2$, $P$ is the Kane matrix element, $E_g$ is the band gap. Eq. (\ref{Rash})  results in $k^3$ 2D terms, which will become linear in $k$ upon quantization in a 1D wire.
 We neglect terms $\propto J_i^3$ leading to the 2D linear in $k$ terms, as they come from admixtures of distant bands and are much smaller than (\ref{Rash}) \cite{Pikus,Knap,Ivchenko}. We also omit the effect of magnetic field in the Rashba and Dresselhaus terms. 

The effective 2D Hamiltonian $H_2$ is given by
\begin{eqnarray}
\label{e8}
&H_2=\frac{p^2}{2m}+
V_{\mathbf{r}}+\tilde{\alpha}\sigma_z[\nabla_{\mathbf{r}}\tilde{V}_\mathbf{r}\times\mathbf{p}]_z
+u(\sigma_xp_x+\sigma_yp_y)\nonumber+\\
&\sum_n\beta_n p^3(\sigma_x\sin{n\phi}-(n-2)\sigma_y\cos{n\phi})+g\mu_B B \sigma_z,
\end{eqnarray}
$n=1,3$. The effective 2D mass $m$  and $g-$factor are 
\begin{eqnarray}
\label{e9}
&\frac{m_0}{m}=&\gamma_1+\gamma_2 -3a^2\gamma_2+3a^2(\gamma_1^2-4\gamma_2^2)^{1/2}f, \\
\label{e20}
&g=&6\kappa+\frac{27}{2}q-6a^2\gamma_2+6a^2(\gamma_1^2-4\gamma_2^2)^{1/2}f,
\end{eqnarray}
where $a=\frac{\gamma_3}{\gamma_2}$, $f=\cot{\frac{\pi}{2}\sqrt{\frac{\gamma_1-2\gamma_2}{\gamma_1+2\gamma_2}}}$. 
Similarity of $m_0/m$ and $g$ is due to contributions of terms $\gamma_3\hat{p_z}(p_x[J_zJ_x] + p_y[J_zJ_y])$ reflecting mutual transformation of holes,  into symmetric state energy $\propto p_-p_+= p_x^2 + p_y^2 + i(p_xp_y-p_yp_x)$, and antisymmetric state energy
$\propto p_+p_-= p_x^2 + p_y^2 - i(p_xp_y-p_yp_x)$. In magnetic field, $ i(p_xp_y-p_yp_x) \rightarrow \frac{\hbar eH}{c}$, making the two last terms in (\ref{e20}) twice those in (\ref{e9}). This simple picture confirms results obtained by considering Landau quantization of holes \cite{simion2014magnetic}.

There are three spin-orbit terms in the ground state Hamiltonian (\ref{e8}) strongly affected by mutual transformation of holes.
The Dresselhaus term is given by
\begin{equation}
u=\frac{1}{2}\left(\frac{\pi}{d}\right)^2\delta\alpha_v\left[1- a\left(1-\sqrt{\nu}\frac{\gamma_1}{\gamma_2}f\right)\right],
\end{equation} where $a=\frac{\gamma_3}{\gamma_2}$. The Rashba term is defined by $\beta_1$, $\beta_3$:
\begin{equation}
\label{2DR}
\beta_n=\frac{3eFd^4}{4\hbar^3\pi^4}\left[A_n\left(\frac{4f/\sqrt{\nu}}{1-\nu}-\frac{3+\nu}{4\nu}\right)+\tilde{\zeta}_n\right].
\end{equation}
Here $A_n=a(a+n-2)$, and the asymmetric part of the potential $eFz$ for $-\frac{d}{2}<z<\frac{d}{2}$, due to doping and external electric field, is assumed to be the same for valence and conduction electrons. 
Distinct offset potentials, $U_c$ for electrons and $U_v$ for holes, result in nonzero $\tilde{\zeta}_n=\pi^2\hbar\zeta (n-1) (U_c-U_v)/(2d^2U_v)$. For the ground state, $\beta_1$ and $\beta_3$ describe $p^3$ Rashba coupling. 
The $\beta_1$-term has no $\zeta$-contribution and vanishes in spherical approximation $a=1$. In our approach, the first term in square brackets in (\ref{2DR}) arises from the matrix element $\langle \pm|eFz|\mp\rangle$, and accounts for an infinite number of terms of the same order that arise in a perturbative approach. Perturbative consideration\cite{Winkler} accounted just for two excited states and resulted only in $(n=3)$-type terms. 

The $\mathbf{\sigma_z}$-term in Eq.(\ref{e8}) results in skew scattering off 
impurities in the 2D hole gas, but our interest here is rather such term  due to the potential confining the wire.
The effective potentials  in Eq. (\ref{e8}) are defined by
\begin{equation}
\tilde{V}_{\mathbf{r}}=\frac{3}{2d}\int_{-d/2}^{d/2}\left[U_v(z,\mathbf{r})S_z^2+ \frac{\zeta\pi^2\hbar}{d^2}U_e(z,\mathbf{r})C_z^2\right] dz,
\end{equation}
$V_{\mathbf{r}}=\frac{2}{d}\int_{-d/2}^{d/2}U_v(\mathbf{r},z)C_z^2dz$, and the constant $\tilde{\alpha}=d^2/\pi^2\hbar$.

Reducing 2D Hamiltonian (\ref{e8}) to the 1D Hamiltonian for the wire, we arrive to our central result
\begin{eqnarray}
\label{e10}
H_1=\frac{p_y^2}{2m}+\alpha\sigma_zp_y+u\sigma_yp_y+\beta\sigma_xp_y+M_z\sigma_z,
\end{eqnarray}
where $M_z=g\mu_B B/2$, and $\beta=(\beta_1+3\beta_3)(\pi\hbar/w)^2$. We neglect a small change in 1D mass $\propto d^2/w^2$ due to quartic terms 
$\propto p_x^2p_y^2 (d/\hbar)^2$, which are omitted in the 2D Hamiltonian (\ref{e8}). In order to find $\alpha$, we consider a model, in which potentials 
$U_{v,e}(z,\mathbf{r})$ are products of functions that depend only on $z$ and only on $x$.  We take symmetric parts of the wire potential for holes and electrons $U^{(s)}_v(x)=U_v^{(w)}\theta(x)$ and  $U^{(s)}_e(x)=U_e^{(w)}\theta(x)$, correspondingly,  where
$\theta (x)=-1$ at $-w/2<x<w/2$ and  $\theta (x)=0$ otherwise. 
Then in the presence of the electric field $F_x$ acting on both electrons and holes we obtain $\alpha=3\zeta eF_x(1-U_e^{(w)}/U_v^{(w)})$. 
The results for the constant $\alpha$, which turns out to be independent of $\tilde{V}$ , and for the contribution to $\beta_3$ arising from 
(\ref{Rash}) are the consequences of the Ehrenfest theorem on vanishing average gradient of a potential in confined states.  

The ground state in a wire is separated by a sizable gap from excited states for in-plane and $z-$direction quantization.
 Eq. (\ref{e10}) fully takes into account physics and all features of charge carrier holes via the modified mass, 
g-factor and spin-orbit constants. The advantage of this electron-like Hamiltonian is that methods similar to those in \cite{lutchyn2011search,sau2010non,stanescu2011majorana} 
can be used investigate the Majorana modes. 


 \paragraph{The existence of Majorana fermions.}
After two rotations in spin space with respect to z and x axis, (\ref{e10}) can be transformed into the equivalent sinlgle-hole Hamiltonian ( index $y$ for p is dropped for brevity):
\begin{equation}
\label{e11}
H_1=\frac{p^2}{2m}+\gamma\sigma_yp+ M_z(cos\theta\sigma_z+sin\theta\sigma_y)
\end{equation}
where $\gamma=\sqrt{\alpha^2+\beta^2+u^2}$ describes the total spin-orbit interaction and $sin\theta=\frac{\alpha}{\sqrt{\alpha^2+\beta^2+u^2}}$ 
measures the alignment of the Zeeman field with the spin-orbit field. The spin wavefunctions in the ground state of (\ref{e11}) are labeled  + and -.
The superconducting pairing can arise due to the proximity effect \cite{Rokhinson2,mourik2012signatures,Tinkham,Kouwenhoven,Marcus,Yacoby,Nadgorny}. 
Assuming holes only in the ground band of the wire, we write the superconducting Hamiltonian as
$
H_{SC}=\int d\mathbf{r}{\Delta e^{i\phi}\hat{c}^\dagger_+\hat{c}^\dagger_-+H.c.},
$
where $\hat{c}^\dagger_\pm$ are the creation operators that add holes to + and -  states in the ground subband, and $\Delta e^{i\phi}$ is the pairing potential. The corresponding BdG Hamiltonian in the Nambu spinor $\mathbf{\tau}$ space reads:
\begin{eqnarray}
\label{e13}
H_{BdG}&=&(\frac{p^2}{2m}-\mu+\gamma\sigma_yp)\tau_z+M_z(cos\theta\sigma_z+sin\theta\sigma_y)\nonumber\\
&+&\Delta cos\phi\tau_x-\Delta sin\phi\tau_y.
\end{eqnarray}
In order to show the existence of Majorana modes, we prove that there is a non-degenerate solution at $E=0$ of the BdG equation $H_{BdG}\Psi=E\Psi$ \cite{sau2010non}. 
Due to the particle-hole symmetry, this solution should take the form $\Psi=(\psi,i\sigma_y\psi^*)$,
where $\psi$ is a two-spinor. We have
\begin{eqnarray}
\label{e16}
[(\frac{p^2}{2m}-\mu+\gamma\sigma_yp)+M_z(cos\theta\sigma_z+sin\theta\sigma_y)]\psi\nonumber\\
+\Delta e^{i\phi}i\sigma_y\psi^*=0
\end{eqnarray}
Writing $\psi=\psi_R+i\psi_I$ and using Eq. (\ref{e16}), we obtain
\begin{eqnarray}
\label{e17}
(\frac{p^2}{2m}-\mu+\gamma\sigma_yp+M_zcos\theta\sigma_z+\Delta cos\phi i\sigma_y)\psi_R\nonumber\\
+(M_zsin\theta+\Delta sin\phi)i\sigma_y\psi_I=0
\end{eqnarray}
\begin{eqnarray}
\label{e18}
(\frac{p^2}{2m}-\mu+\gamma\sigma_yp+M_zcos\theta\sigma_z-\Delta cos\phi i\sigma_y)\psi_I\nonumber\\
-(M_zsin\theta-\Delta sin\phi)i\sigma_y\psi_R=0
\end{eqnarray}
Analyzing Majorana modes so far, we followed \cite{sau2010non}. However, we observe from our Eq.(\ref{e17}) that the sign of the Zeemann term in the Eq. (59)  of \cite{sau2010non} should be +,  not -. As a result,  Eqs.(\ref{e17}) and (\ref{e18}) for $\Psi_I$ and $\Psi_R$ cannot be both decoupled by a unique choice of the order parameter phase $\phi$, $M_zsin\theta=\Delta sin\phi$. For such choice,  Eq. (\ref{e18}) will "decouple", containing and defining $\Psi_I$ only, but Eq.(\ref{e17}) will still connect $\Psi_I$ and $\Psi_R$. Similarly, for $M_zsin\theta=-\Delta sin\phi$, Eq. (\ref{e17}) will contain and define $\Psi_R$ only, but Eq.(\ref{e18}) will still connect $\Psi_I$ and $\Psi_R$.

\begin{figure}
\centering
\subfigure
{
\begin{minipage}[b]{0.22\textwidth}
\includegraphics[width=4.5cm,height=4cm]{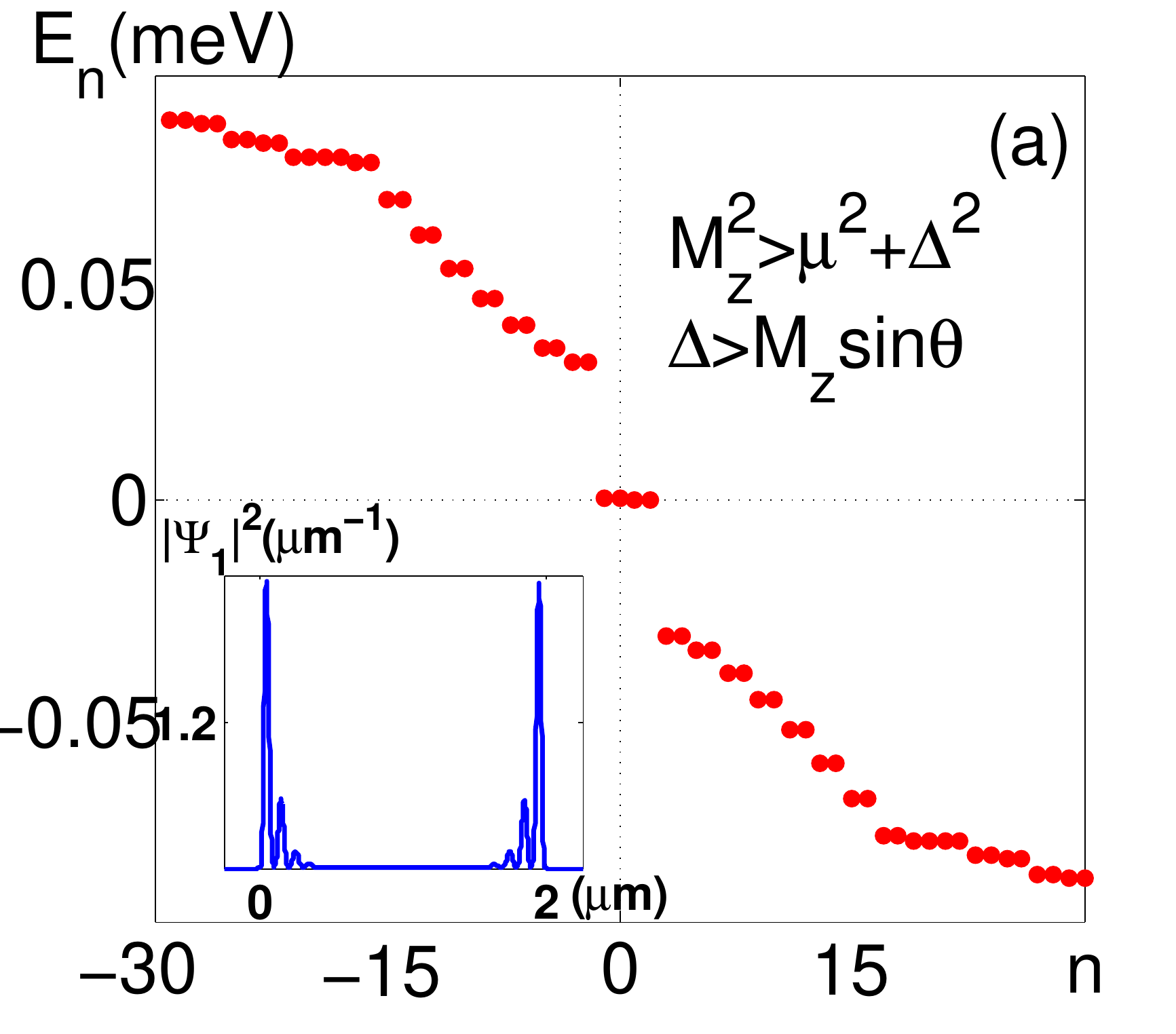}
\end{minipage}
\label{2a}
}
\subfigure
{
\begin{minipage}[b]{0.22\textwidth}
\includegraphics[width=4.5cm,height=4cm]{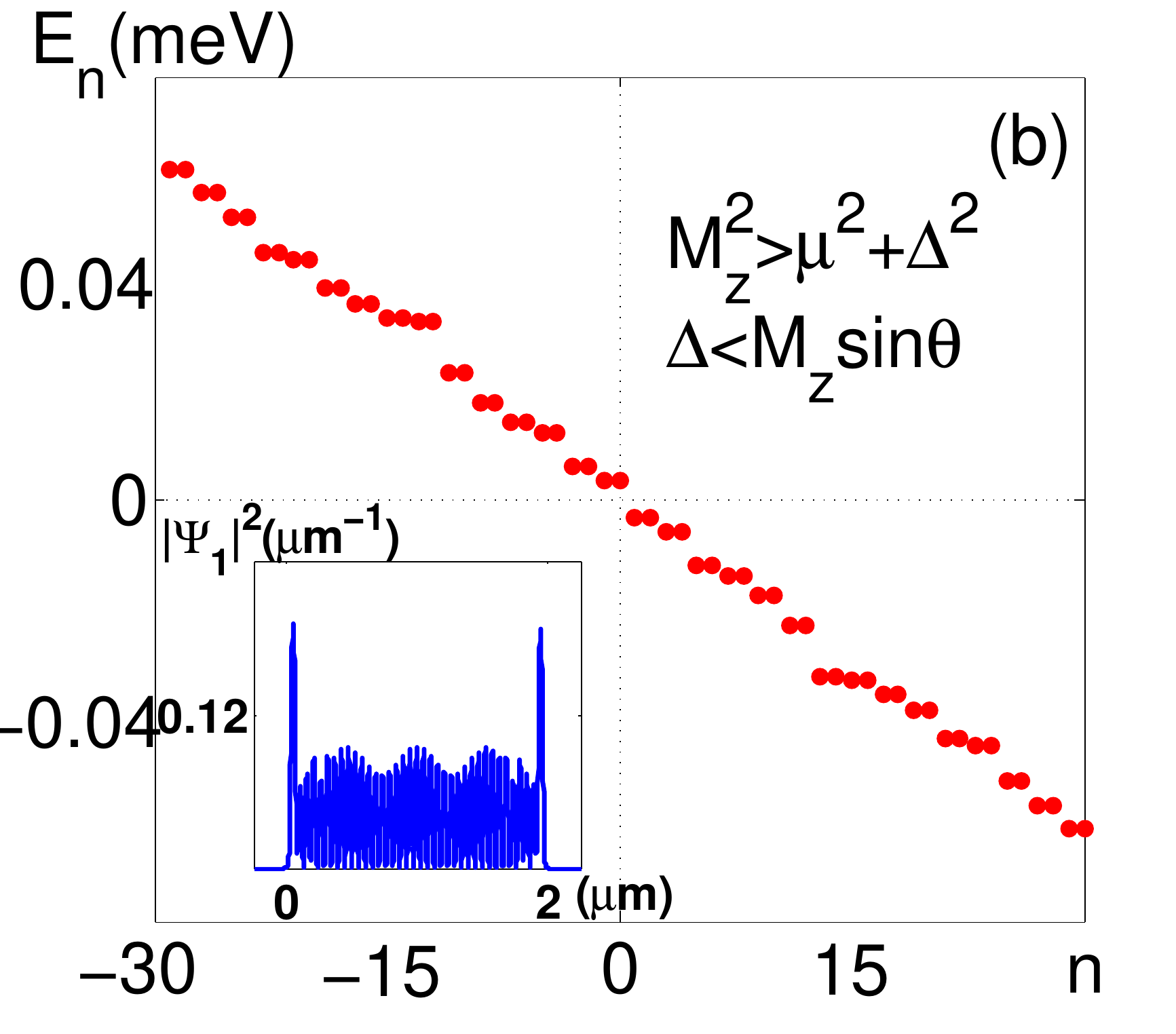}
\end{minipage}
\label{2b}
}
\subfigure
{
\begin{minipage}[b]{0.22\textwidth}
\includegraphics[width=4.5cm,height=4cm]{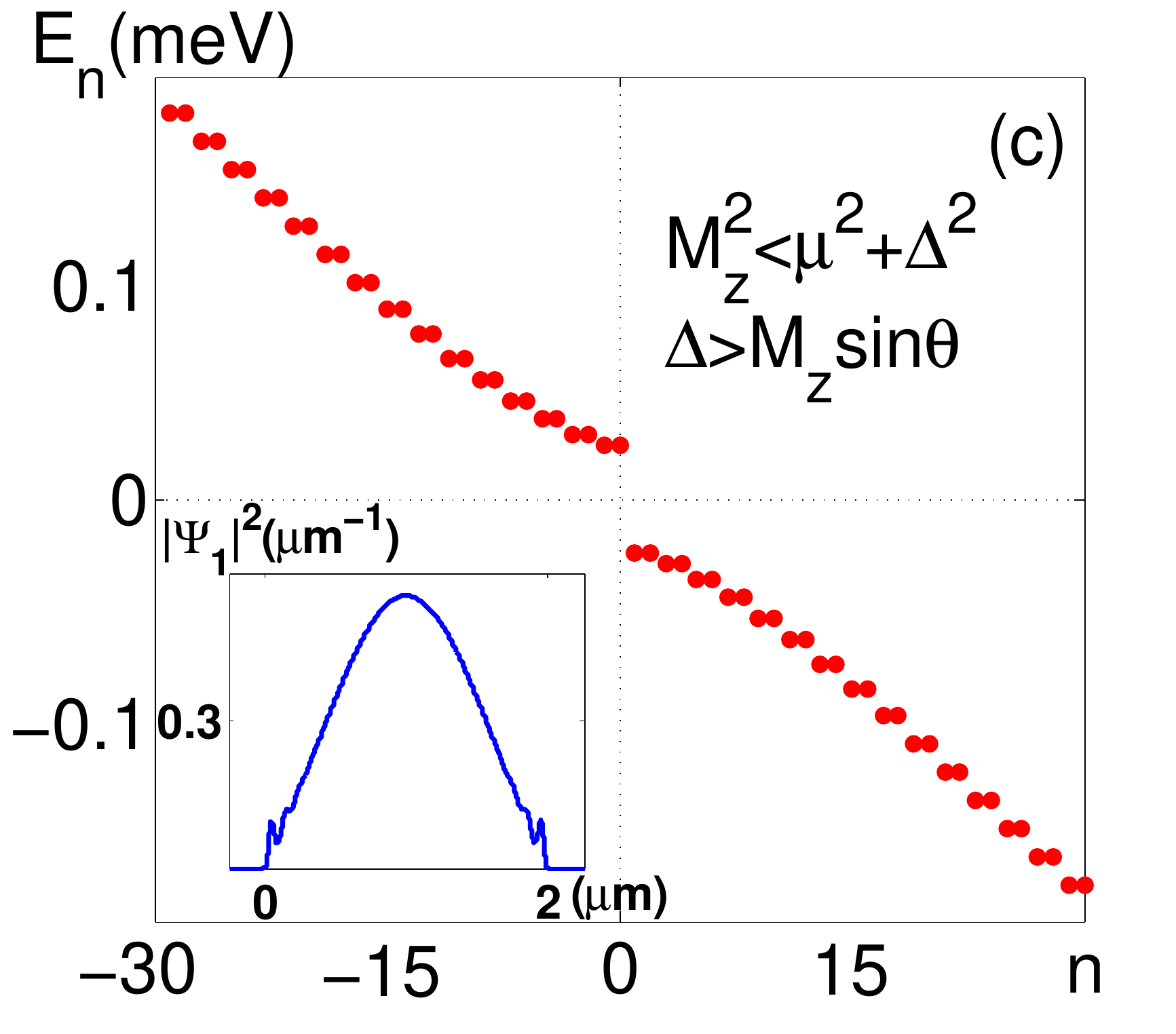}
\end{minipage}
\label{2c}
}
\subfigure
{
\begin{minipage}[b]{0.22\textwidth}
\includegraphics[width=4.5cm,height=4cm]{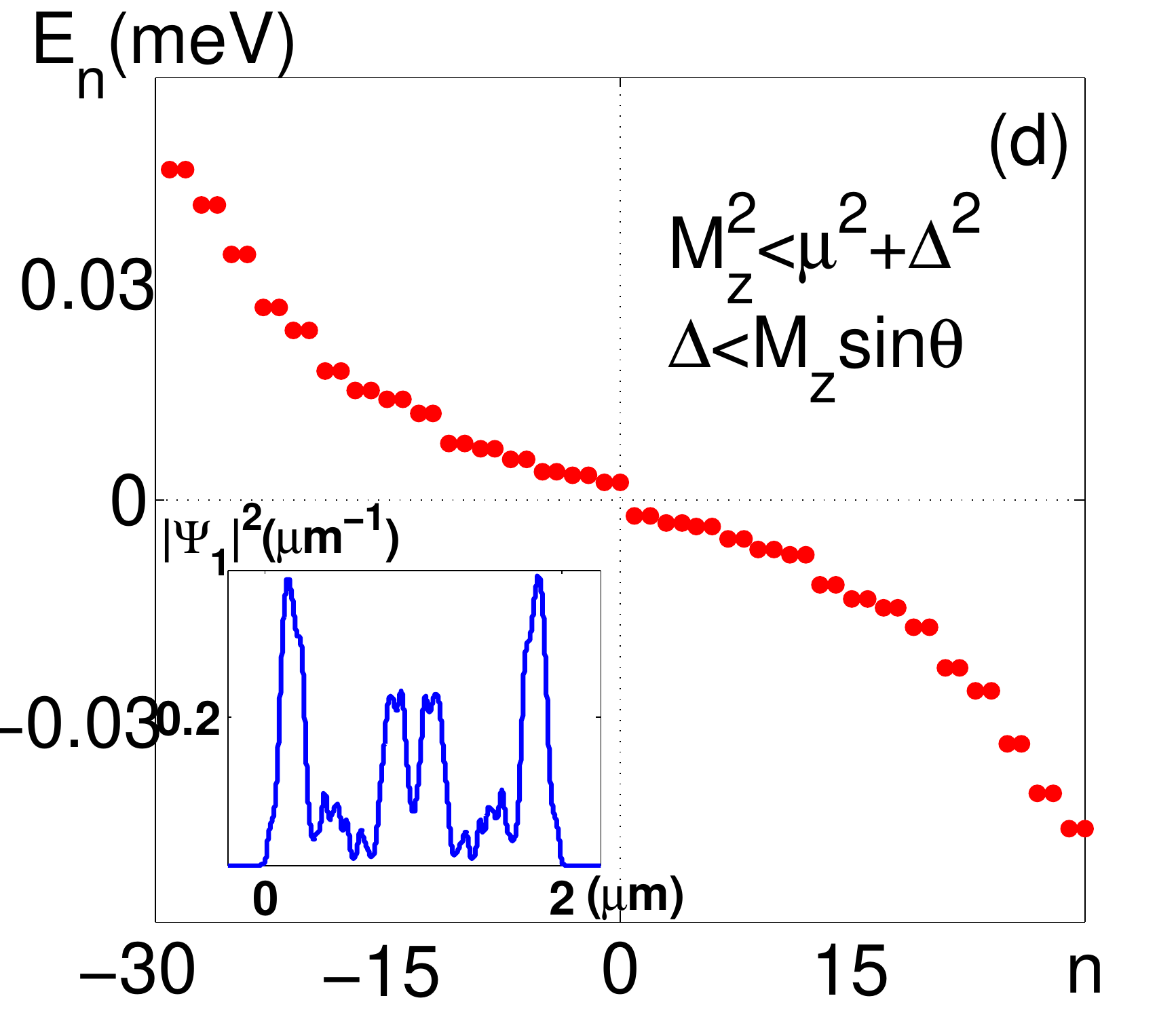}
\end{minipage}
\label{2d}
}
\caption{(Color online) The BdG energy spectra $E_n$ in a 2 $\mu m$ long GaAs quantum wire.  $B=0.8T$,  $E_{so}=0.2 meV$. a: $\mu=0.5M_z$,$\Delta=0.6M_z$, $sin\theta=0$. The zero energy solution exists and is well separated by a gap from the excited states. Inset: Majorana zero mode is localized at the boundary of the quantum wire. b,c,d: Parameters: b - $\mu=0.5M_z$,$\Delta=0.6M_z$, $sin\theta=0.7$; c - $\mu=0.5M_z$,$\Delta=M_z$, $sin\theta=0$; d - $\mu=0.9M_z$,$\Delta=0.6M_z$, $sin\theta=0.7$. There are no zero energy solutions. Insets: Wavefunctions of the lowest-lying states. Majorana zero modes disappear in cases b,c,d.  \label{fg2}}
\end{figure}

We now demonstrate that Majorana zero mode exists for any of the choices  $M_zsin\theta=\lambda\Delta sin\phi$, where $\lambda =\pm 1$. 
The equations for the corresponding functions $\psi_{R}$ $(\lambda =1)$ or $\psi_{I}$ $(\lambda =-1)$  are given by 
 ($\lambda=\pm1$):
\begin{equation}
\label{e19}
\left(\begin{array}{cc}
\frac{p^2}{2m}-\mu+M_zcos\theta & \lambda\Delta cos\phi-i\gamma p\\
-\lambda\Delta cos\phi+i\gamma p & \frac{p^2}{2m}-\mu-M_zcos\theta
\end{array}\right)\psi_{R/I}=0.
\end{equation}

\begin{figure}
\centering
\subfigure 
{
\begin{minipage}[b]{0.22\textwidth}
\includegraphics[width=4.5cm,height=4cm]{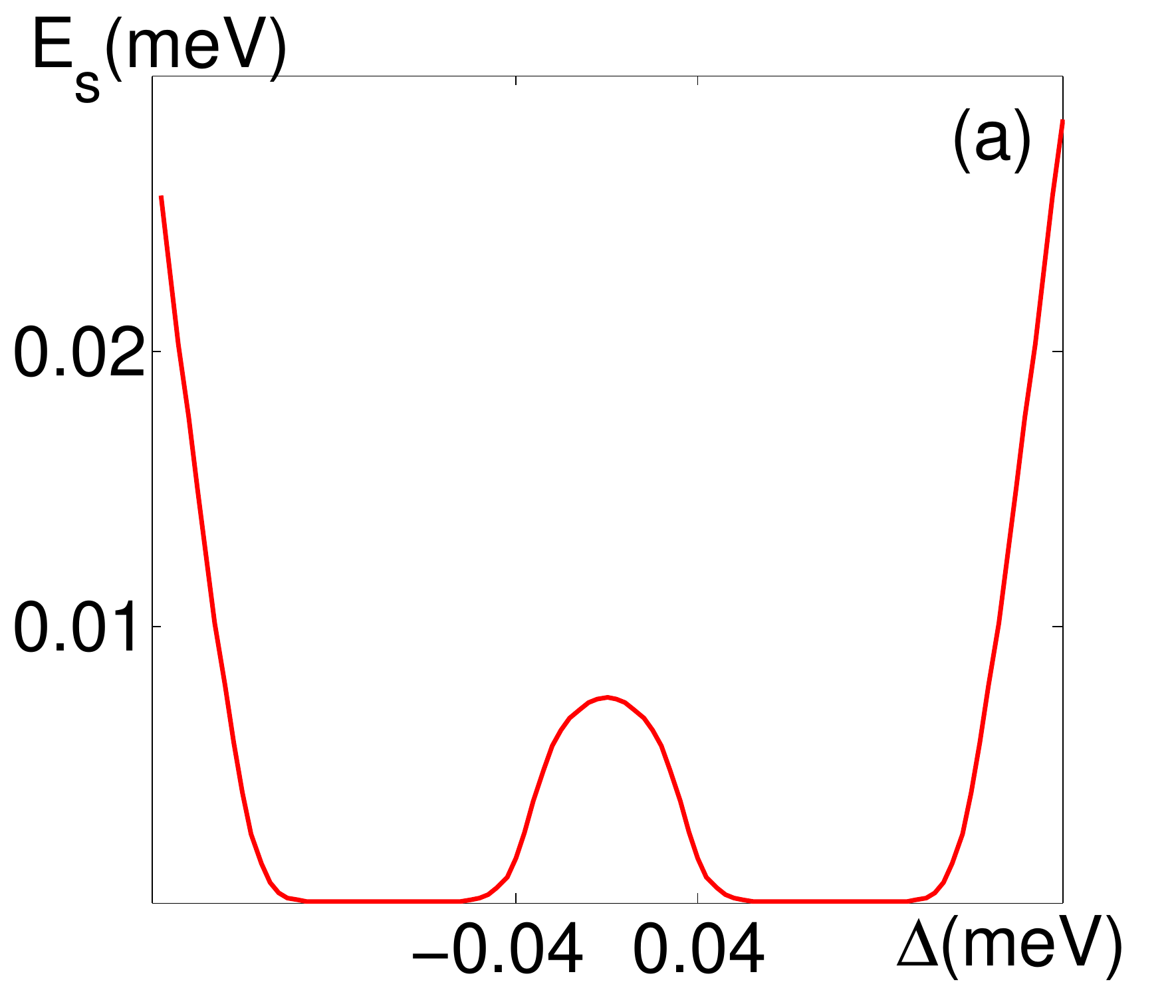}
\end{minipage}
\label{3a}
}
\subfigure
{
\begin{minipage}[b]{0.22\textwidth}
\includegraphics[width=4.5cm,height=4cm]{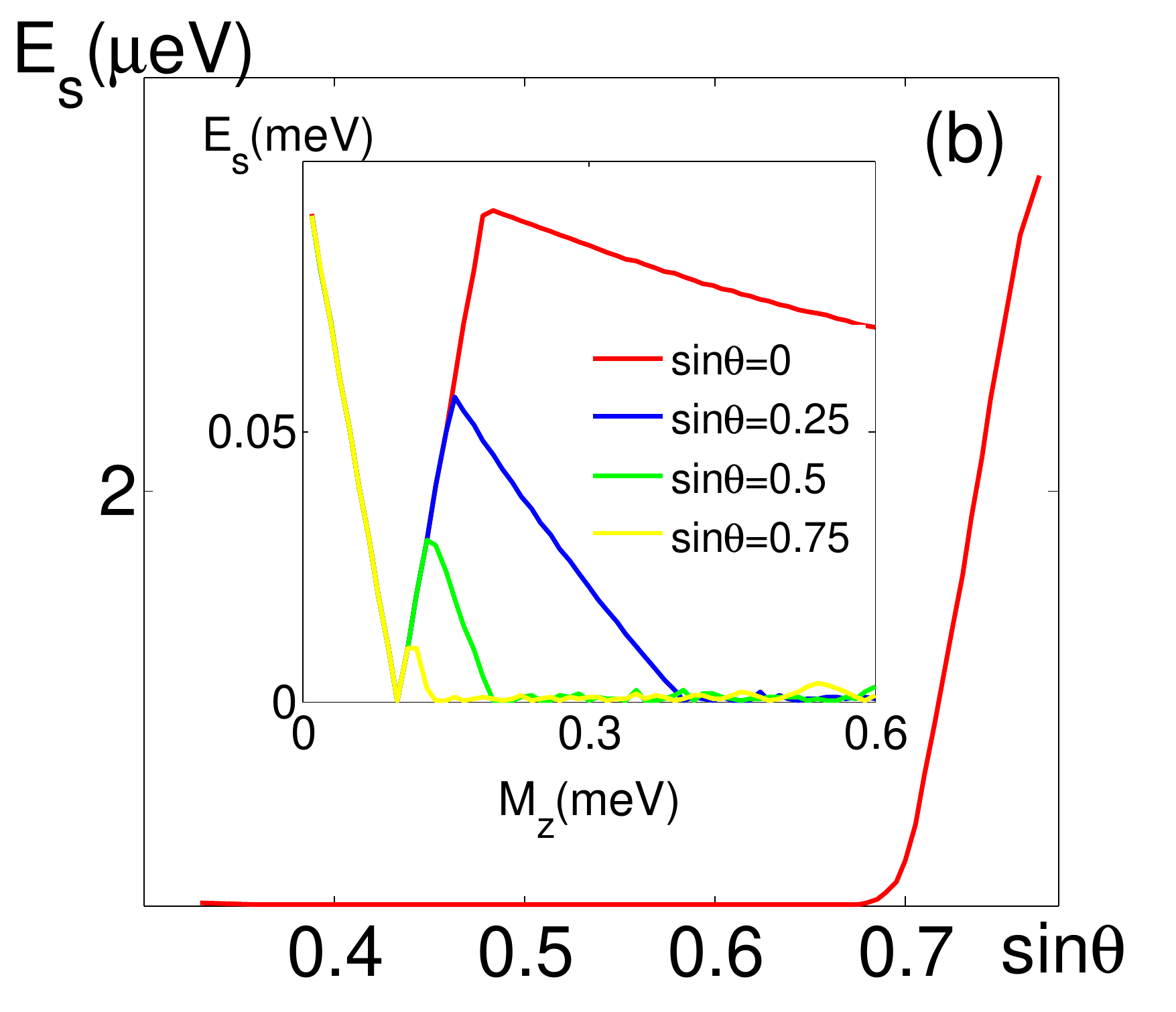}
\end{minipage}
\label{3b}
}

\caption{(Color online) a: the BdG ground state energy $E_s$ for different $\Delta$. $M_z=0.2 meV$, $E_{so}=0.2 meV$, $\mu=0.6M_z$ and $sin\theta=0.2$.
$E_s\ne 0$ at $|\Delta|<0.2M_z$ or $|\Delta|>0.8M_z$, which correspond to $\Delta<M_zsin\theta$ and $M_z^2<\mu^2+\Delta^2$, respectively. b: $E_s$ for different $sin\theta$ ($\mu=0.5M_z$, $\Delta=0.7M_z$). $E_s$ is non-zero at $\Delta<M_zsin\theta$. Inset: Excitation gap versus $M_z$. $\Delta=0.1 meV$, $\mu=0$. The gap closes at $M_z=\Delta$ or $\Delta/M_z=sin\theta$.\label{fg3}}
\end{figure}

Assumption $\psi_{I/R}\sim e^{-\tau y}$ leads to the secular equation
\begin{equation}
\label{e30}
\frac{\tau^4}{4m^2}+(\frac{\mu}{m}+\gamma^2)\tau^2+2\lambda\gamma\Delta cos\phi\tau+C_0=0,
\end{equation}
where $C_0=\mu^2+\Delta^2-M_z^2$. Using Vieta's formulas, for $C_0<0$ we find 3 roots with $Re[\tau]>0$ and 1 root with $Re[\tau]<0$ for $\lambda=-1$, or 
 1 root with $Re[\tau]>0$ and 3 roots with $Re[\tau]<0$ for $\lambda=1$.
For $C_0>0$ there are two roots with $Re[\tau]\lessgtr0$. Because there are one normalization and four boundary conditions, 
a unique bound state solution exists on the boundary between the topological non-trivial 
phase with $C_0=\Delta^2+\mu^2-M_z^2<0$ and a trivial phase with $C_0>0$, in much the same way as in \cite{sau2010generic,lutchyn2010majorana}. 
Once $\psi_{R/I}$ is found from Eq. (\ref{e17}) or Eq. (\ref{e18}) for a given choice of $\phi$, $\psi_{I/R}$ can be found using the other equation coupling  $\psi_{R/I}$ and $\psi_{I/R}$. Normalization of wavefunctions is achieved using equations that contain only $\psi_{R}$ or only  $\psi_{I}$. At $E=0$ these $\psi_{I/R}$ define the wavefiunction of the Majorana bound state. Then the criterion for the topological superconductivity in hole wires is
\begin{equation}
\label{e21}
M_z^2>\mu^2+\Delta^2.
\end{equation}
In order to arrive to this result, we used one of the relations $M_zsin\theta=\pm\Delta sin\phi$.
As $\phi$ can be made arbitrary,  an extra constraint for existence of Majorana modes is
\begin{equation}
\label{constraint}
|\Delta|\ge |M_zsin\theta|.
\end{equation}

Does Majorana mode exist when none of the equations (\ref{e17}) or (\ref{e18}) can be decoupled? 
We now show that for arbitrary $\phi$ the constraint $|\Delta|> |M_zsin\theta|$ is necessary for existence of a topological superconductor in proximitized hole wires.
If $sin\theta=1$, i.e. the Zeeman field is aligned with the spin-orbit field, the Majorana modes do not exist. If $sin\theta=0$, Majorana modes arise if Eq. (\ref{e21}) is satisfied.
 Furthermore, Majorana modes exist at $M_zsin\theta=\pm\Delta sin\phi$, for " intermediate"  $\theta$. This precludes the possibility that only $sin\theta=0$ case, i.e. when the BdG equations are equivalent to those with real coefficients, gives Majorana modes, while for all other $\theta$ these modes disappear.
Thus, there is a critical angle $\theta_c\ne 0$ that separates regions of the phase diagram  where the topological superconductivity exists. Our result is that this critical angle is given by $\Delta=M_zsin\theta_c$. 
To see this, we solve the BdG equation numerically and find the energy spectra and the wavefunctions, Fig.\ref{fg2}. We observe that the Majorana zero mode exists only when both conditions are satisfied, and disappears when one of the conditions is not fullfilled. The extra constraint means that the $\mathbf\sigma_z$-term should not be too large compared to  the Rashba and Dresselhaus terms. In Fig.\ref{fg3} we show 
the ground state BdG energies for different $\Delta$ and $sin\theta$.

We suggest that hole wires in GaAs or InSb structures, in which surface levels are in the forbidden gap and make 
proximity effect in p-type structures feasible,  are favorable for detection of Majorana modes.
For $w=80 nm$ nanowire, lithographically developed from an unstrained $d=20nm$ quantum well in AlGaAs/GaAs/AlGaAs Carbon-doped heterostructure grown along [001],  at $\gamma_1=6.8$, 
$\gamma_2=2.1$, $\gamma_3=2.9$, $\kappa=1.2$, $q=0.04$, $P=10 eV \AA$, $E_g= 1.52 eV$ and \cite{Ivchenko} $\delta\alpha_v=76.7$ eV $\AA^3$, assuming $eF_z=2\times 10^4 V/cm,$ we obtain $m=0.15$ m$_0$ ( $m=0.25 m_0$ when adjustments are made 
for the effects of finite depth of the well and one-sided doping), $g=5$, and $\gamma= 70$ meV $\AA$. At 2D density $n_s = 2\times10^{10} cm^{-2}$, $\mu =0.14 meV$, and holes are only in the ground subband in the wire.
For superconductivity caused by proximity to  NbN, $\Delta \sim$ $0.15meV$, and the transition between topological and non-topological superconducting order occurs 
at $H\sim 0.8 T$. Then $E_{so}\sim M_z\sim 0.2 meV$, and the lowest single-hole state is in transition from type Fig1.b to a type Fig1c, i.e., spin-orbit coupling is somewhat strong and leads to effective p-pairing. 

For a similar InSb wire in InSb/AlInSb structure, at $\gamma_1=40.1$, $\gamma_2=18.1$, $\gamma_3=19.2$, $\kappa=17.0$, $q=0.5$, $P=9.6 eV \AA$, $E_g= 0.23 eV$ and \cite{Pikus} $\delta\alpha_v=70$ eV $\AA^3$,
we obtain $\gamma\sim 250 meV $\AA, and $g\sim 90$. Due to strain in this system, we take the wire mass $m=0.04m_0$ \cite{Santos} [according to (\ref{e9}) at zero strain, $m=0.018m_0$].
Then $E_{so}\sim 0.4 meV$. In proximity with NbN at $\Delta \sim$ $0.15meV$,
 $n_s = 2\times 10^{10} cm^{-2}$, we  obtain that the transition between topological and non-topological superconductivity in the InSb hole wire at $H\sim 0.4 T$. In this case $ M_z> E_{so}$, and the ground state of (\ref{e11}) has single-minimum Fig.1c shape, which also takes place in an electron InSb wire \cite{mourik2012signatures} with $E_{so}$ 4 times smaller than here. 
 
{\it Conclusion}. We considered charge carrier hole wires non-perturbatively, including the effect of mutual transformation of heavy and light holes upon reflection from heteroboundaries,  and derived the hole g-factor and spin-orbit interactions. 
Beyond Majorana context, these results are important for the field of spin-based electronics, and relevant for gate-induced, lithographically defined and cleaved edge overgrowth \cite{Hughes} wires, 
 for generation, manipulation and transmission of spin currents. 
We considered a criterion for transition from non-topological to topological superconducting order in proximity-coupled hole nanowires, showing that Majorana modes can arise even if the
Bogoliubov-De Gennes equations for real and imaginary components of the wavefunction cannot be decoupled. 
We evaluated Majorana settings in GaAs and InSb hole wires, and demonstrated that these systems potentially exhibit considerably stronger p-type proximity-induced superconducting pairing compared to 
InSb electron system in similar magnetic fields. 

Authors thank J. Shabani, G. Simion, M. Radu, L. Rokhinson and Y. Shu for helpful discussions. This work is supported by the U.S. Department of Energy, Office of Basic Energy Sciences, Division of Materials Sciences and Engineering under Award DE-SC0010544.


\end{document}